\documentclass[sigconf]{acmart}

\usepackage{subfigure}
\usepackage{multirow}
\usepackage{color}
\usepackage{adjustbox}
\usepackage{url}
\usepackage{bbding}
\usepackage{balance}
\newcommand{\etal}{\emph{et al.}~}
\newcommand{\ie}{\emph{i.e.}}
\newcommand{\eg}{\emph{e.g.}}

\def\BibTeX{{\rm B\kern-.05em{\sc i\kern-.025em b}\kern-.08emT\kern-.1667em\lower.7ex\hbox{E}\kern-.125emX}}
    
\copyrightyear{2019}
\acmYear{2019}
\setcopyright{acmcopyright}
\acmConference[MM '19] {Proceedings of the 27th ACM International Conference on Multimedia}{October 21--25, 2019}{Nice, France}
\acmBooktitle{Proceedings of the 27th ACM International Conference on Multimedia (MM'19), October 21--25, 2019, Nice, France}
\acmPrice{15.00}
\acmDOI{10.1145/3343031.3351084}
\acmISBN{978-1-4503-6889-6/19/10}

\begin{document}

\fancyhead{}

\title[Lightweight Image SR with IMDN]{Lightweight Image Super-Resolution with Information Multi-distillation Network}

\author{Zheng Hui}
\affiliation{%
	\institution{School of Electronic Engineering, Xidian University}
	\streetaddress{No.2, South Taibai Road}
	\city{Xi'an}
	\country{China}
}
\email{zheng\_hui@aliyun.com}

\author{Xinbo Gao}
\affiliation{%
	\institution{School of Electronic Engineering, Xidian University}
	\streetaddress{No.2, South Taibai Road}
	\city{Xi'an}
	\country{China}
}
\email{xbgao@mail.xidian.edu.cn}

\author{Yunchu Yang}
\affiliation{%
	\institution{School of Electronic Engineering, Xidian University}
	\streetaddress{No.2, South Taibai Road}
	\city{Xi'an}
	\country{China}
}
\email{yc\_yang@aliyun.com}

\author{Xiumei Wang}
\authornote{Corresponding author}
\affiliation{%
	\institution{School of Electronic Engineering, Xidian University}
	\streetaddress{No.2, South Taibai Road}
	\city{Xi'an}
	\country{China}
}
\email{wangxm@xidian.edu.cn}

%
\renewcommand{\shortauthors}{Hui and Gao, et al.}

%
\begin{abstract}
In recent years, single image super-resolution (SISR) methods using deep convolution neural network (CNN) have achieved impressive results. Thanks to the powerful representation capabilities of the deep networks, numerous previous ways can learn the complex non-linear mapping between low-resolution (LR) image patches and their high-resolution (HR) versions. However, excessive convolutions will limit the application of super-resolution technology in low computing power devices. Besides, super-resolution of any arbitrary scale factor is a critical issue in practical applications, which has not been well solved in the previous approaches. To address these issues, we propose a lightweight information multi-distillation network (IMDN) by constructing the cascaded information multi-distillation blocks (IMDB), which contains distillation and selective fusion parts. Specifically, the distillation module extracts hierarchical features step-by-step, and fusion module aggregates them according to the importance of candidate features, which is evaluated by the proposed contrast-aware channel attention mechanism. To process real images with any sizes, we develop an adaptive cropping strategy (ACS) to super-resolve block-wise image patches using the same well-trained model. Extensive experiments suggest that the proposed method performs favorably against the state-of-the-art SR algorithms in term of visual quality, memory footprint, and inference time. Code is available at \url{https://github.com/Zheng222/IMDN}.
\end{abstract}

%
%

\ccsdesc[500]{Computing methodologies~Computational photography}
\ccsdesc[300]{Computing methodologies~Reconstruction}
\ccsdesc[100]{Computing methodologies~Image processing}

%
\keywords{image super-resolution; lightweight network; information multi-distillation; contrast-aware channel attention; adaptive cropping strategy}

%
\maketitle

\section{Introduction}\label{sec:introcution}
Single image super-resolution (SISR) aims at reconstructing a high-resolution (HR) image from its low-resolution (LR) observation, which is inherently ill-posed because many HR images that can be downsampled to an identical LR image. To address this problem, numerous image SR methods~\cite{VDSR,MemNet,SRDenseNet,RDN,IDN,RCAN} based on deep neural architectures~\cite{VGG,ResNet,DenseNet} have been proposed and shown prominent performance.

Dong~\etal~\cite{SRCNN,SRCNN-Ex} first developed a three-layer network (SRCNN) to establish a direct relationship between LR and HR. Then, Wang~\etal~\cite{CSCN} proposed a neural network according to the conventional sparse coding framework and further designed a progressive upsampling style to produce better SR results at the large scale factor (\eg, $\times 4$). Inspired by VGG model~\cite{VGG} that used for ImageNet classification, Kim~\etal~\cite{VDSR,DRCN} first pushed the depth of SR network to $20$ and their model outperformed SRCNN by a large margin. This indicates a deeper model is instructive to enhance the quality of generated images. To accelerate the training of deep network, the authors introduced global residual learning with a high initial learning rate. At the same time, they also presented a deeply-recursive convolutional network (DRCN), which applied recursive learning to SR problem. This way can significantly reduce the model parameters. Similarly, Tai~\etal proposed two novel networks, and one is a deep recursive residual network (DRRN)~\cite{DRRN}, another is a persistent memory network (MemNet)~\cite{MemNet}. The former mainly utilized recursive learning to reach the goal of economizing parameters. The latter model tackled the long-term dependency problem existed in the previous CNN architecture by several memory blocks that stacked with a densely connected structure~\cite{DenseNet}. However, these two algorithms required a long time and huge graphics memory consumption both in the training and testing phases. The primary reason is the inputs sent to these two models are interpolation version of LR images and the networks have not adopted any downsampling operations. This scheme will bring about a huge computational cost. To increase testing speed and shorten the testing time, Shi~\etal\cite{ESPCN} first performed most of the mappings in low-dimensional space and designed an efficient sub-pixel convolution to upsample the resolutions of feature maps at the end of SR models.

To the same end, Dong~\etal proposed fast SRCNN (FSRCNN)~\cite{FSRCNN}, which employed a learnable upsampling layer (transposed convolution) to accomplish post-upsampling SR. Afterward, Lai~\etal presented the Laplacian pyramid super-resolution network (LapSRN)~\cite{LapSRN} to progressively reconstruct higher-resolution images. Some other work such as MS-LapSRN~\cite{MS-LapSRN} and progressive SR (ProSR)~\cite{ProSR} also adopt this progressive upsampling SR framework and achieve relatively high performance. EDSR~\cite{EDSR} made a significant breakthrough in term of SR performance, which won the competition of NTIRE 2017~\cite{NTIRE2017_dataset,NTIRE2017_methods}. The authors removed some unnecessary modules (\eg, Batch Normalization) of the SRResNet~\cite{SRGAN} to obtain better results. Based on EDSR, Zhang~\etal incorporated densely connected block~\cite{DenseNet,SRDenseNet} into residual block~\cite{ResNet} to construct a residual dense network (RDN). Soon they exploited the residual-in-residual architecture for the very deep model and introduced channel attention mechanism~\cite{SENet} to form the very deep residual attention networks (RCAN)~\cite{RCAN}. More recently, Zhang~\etal also introduced spatial attention (non-local module) into the residual block and then constructed residual non-local attention network (RNAN)~\cite{RNAN} for various image restoration tasks.

The major trend of these algorithms is increasing more convolution layers to improve performance that measured by PSNR and SSIM~\cite{SSIM}. As a result, most of them suffered from large model parameters, huge memory footprints, and slow training and testing speeds. For instance, EDSR~\cite{EDSR} has about $43$M parameters, $69$ layers, and RDN~\cite{RDN} achieved comparable performance, which has about $22$M parameters, over $128$ layers. Another typical network is RCAN~\cite{RCAN}, its depth up to $400$ but the parameters are about $15.59$M. However, these methods are still not suitable for resource-constrained equipment. For the mobile devices, the desired practice should be to pursuing higher SR performance as much as possible when the available memory and inference time are constrained in a certain range. 
Many cases require not only the performance but also high execution speed, such as video applications, edge devices, and smartphones. Accordingly, it is significant to devise a lightweight but efficient model for meeting such demands. 

Concerning the reduction of the parameters, many approaches adopted the recursive manner or parameter sharing strategy, such as~\cite{DRCN,DRRN,MemNet}. Although these methods did reduce the size of the model, they increased the depth or the width of the network to make up for the performance loss caused by the recursive module. This will lead to spending a great lot of calculating time when performing SR processing. To address this issue, the better way is to design the lightweight and efficient network structures that avoid using recursive paradigm. Ahn~\etal developed CARN-M~\cite{CARN} for mobile scenario through a cascading network architecture, but it is at the cost of a substantial reduction on PSNR. Hui~\etal~\cite{IDN} proposed an information distillation network (IDN) that explicitly divided the preceding extracted features into two parts, one was retained and another was further processed. Through this way, IDN achieved good performance at a moderate size. But there is still room for improvement in term of performance.

Another factor that affects the inference speed is the depth of the network. In the testing phase, the previous layer and the next layer have dependencies. Simply, conducting the computation of the current layer must wait for the previous calculation is completed. But multiple convolutional operations at each layer can be processed in parallel. Therefore, the depth of model architecture is an essential factor affecting time performance. This point will be verified in Section~\ref{sec:experiemnts}.

As to solving the different scale factors ($\times 2$, $\times 3$, $\times 4$) SR problem using a single model, previous solutions pretreated an image to the desired size and using the fully convolutional network without any downsampling operations. This way will inevitably lead to a substantial increase in the amount of calculation.  

To address the above issues, we propose a lightweight information multi-distillation network (IMDN) for better balancing performance against applicability. Unlike most previous small parameters models that use recursive structure, we elaborately design an information multi-distillation block (IMDB) inspired by~\cite{IDN}. The proposed IMDB extracts features at a granular level, which retains partial information and further treats other features at each step (layer) as illustrated in Figure~\ref{fig:IMDB}. For aggregating features distilled by all steps, we devise a contrast-aware channel attention layer, specifically related to the low-level vision tasks, to enhance collected various refined information. Concretely, we exploit more useful features (edges, corners, textures,~\etal) for image restoration. In order to handle SR of any arbitrary scale factor with a single model, we need to scale the input image to the target size, and then employ the proposed adaptive cropping strategy (see in Figure~\ref{fig:acs}) to obtain image patches of appropriate size for lightweight SR model with downsampling layers.

The contributions of this paper can be summarized as follows:
\begin{itemize}
	\item We propose a lightweight information multi-distillation network (IMDN) for fast and accurate image super-resolution. Thanks to our information multi-distillation block (IMDB) with contrast-aware attention (CCA) layer, we achieve competitive results with a modest number of parameters (refer to Figure~\ref{fig:parameters}).
	
	\item We propose the adaptive cropping strategy (ACS), which allows the network included downsampling operations (\eg, convolution layer with a stride of 2) to process images of any arbitrary size. By adopting this scheme, the computational cost, memory occupation, and inference time can dramatically reduce in the case of treating indefinite magnification SR.
	
	\item We explore factors affecting actual inference time through experiments and find the depth of the network is related to the execution speed. It can be a guideline for guiding a lightweight network design. And our model achieves an excellent balance among visual quality, inference speed, and memory occupation.
\end{itemize}
\section{Related Work}\label{sec:related-work}
\subsection{Single image super-resolution}
With the rapid development of deep learning, numerous methods based on convolutional neural network (CNN) have been the mainstream in SISR. The pioneering work of SR is proposed by Dong~\etal~\cite{SRCNN,SRCNN-Ex} named SRCNN. The SRCNN upscaled the LR image with bicubic interpolation before feeding into the network, which would cause substantial unnecessary computational cost. To address this issue, the authors removed this pre-processing and upscaled the image at the end of the net to reduce the computation in~\cite{FSRCNN}. Lim~\etal~\cite{EDSR} modified SRResNet~\cite{SRGAN} to construct a more in-depth and broader residual network denoted as EDSR. With the smart topology structure and a significantly large number of learnable parameters, EDSR dramatically advanced the SR performance. Zhang~\etal~\cite{RDN} introduced channel attention~\cite{SENet} into the residual block to further boost very deep network (more than $400$ layers without considering the depth of channel attention modules). Liu~\cite{NLRN} explored the effectiveness of non-local module applied to image restoration. Similarly, Zhang~\etal~\cite{RNAN} utilized non-local attention to better guide feature extraction in their trunk branch for reaching better performance. Very recently, Li~\etal~\cite{feedbackSR} exploited feedback mechanism that enhancing low-level representation with high-level ones. 

For lightweight networks, Hui~\etal~\cite{IDN} developed the information distillation network for better exploiting hierarchical features by separation processing of the current feature maps. And Ahn~\cite{CARN} designed an architecture that implemented a cascading mechanism on a residual network to boost the performance.
\subsection{Attention model}
Attention model, aiming at concentrating on more useful information in features, has been widely used in various computer vision tasks. Hu~\etal~\cite{SENet} introduced squeeze-and-excitation (SE) block that models channel-wise relationships in a computationally efficient manner and enhances the representational ability of the network, showing its effectiveness on image classification. CBAM~\cite{CBAM} modified the SE block to exploit both spatial and channel-wise attention. Wang~\etal~\cite{non-local} proposed the non-local module to generate the wide attention map by calculating the correlation matrix between each spatial point in the feature map, then the attention map guided dense contextual information aggregation.

\section{Method}
\begin{figure*}[htpb]
	\centering
	\subfigure[IMDN]{\label{fig:IMDN}
		\includegraphics[width=0.45\textwidth]{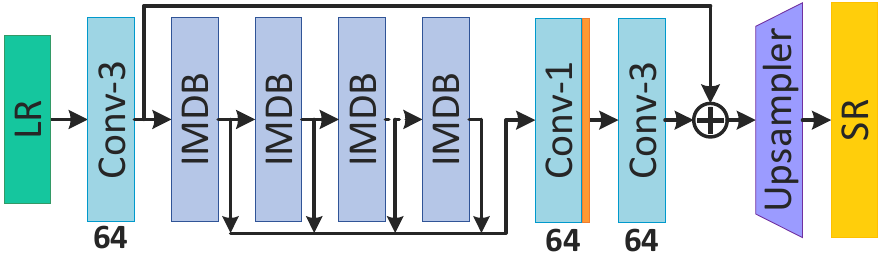}}
	\hfil
	\subfigure[Upsampler]{\label{fig:Upsampler}
		\includegraphics[width=0.08\textwidth]{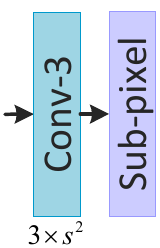}}
	\vspace{-3mm}
	\caption{The architecture of information multi-distillation network (IMDN). (a) The \textcolor{orange}{orange box} represents Leaky ReLU activation function and the details of IMDB is shown in Figure~\ref{fig:IMDB}. (b) s represents the upscale factor.}
\end{figure*}

\subsection{Framework}
In this section, we describe our proposed information multi-distillation network (IMDN) in detail, its graphical depiction is shown in Figure~\ref{fig:IMDN}. The upsampler (see Figure~\ref{fig:Upsampler}) includes one $3 \times 3$ convolution with $3 \times {s^2}$ output channels and a sub-pixel convolution. Given an input LR image ${\mathbf{I}^{LR}}$, its corresponding target HR image ${\mathbf{I}^{HR}}$. The super-resolved image ${\mathbf{I}^{SR}}$ can be generated by
\begin{equation}
{\mathbf{I}^{SR}} = {H_{IMDN}}\left( {{\mathbf{I}^{LR}}} \right),
\end{equation}
where ${H_{IMDN}}\left(  \cdot  \right)$ is our IMDN. It is optimized with mean absolute error (MAE) loss followed most of previous works~\cite{EDSR,IDN,RDN,RCAN,CARN}. Given a training set $\left\{ {\mathbf{I}_i^{LR},\mathbf{I}_i^{HR}} \right\}_{i = 1}^N$ that has $N$ LR-HR pairs. Thus, the loss function of our IMDN can be expressed by
\begin{equation}
{\mathcal L}\left( \Theta  \right) = \frac{1}{N}\sum\limits_{i = 1}^N {{{\left\| {{H_{IMDN}}\left( {I_i^{LR}} \right) - I_i^{HR}} \right\|}_1}} ,
\end{equation}
where $\Theta$ indicates the updateable parameters of our model and ${{{\left\|  \cdot  \right\|}_1}}$ is ${l_1}$ norm. Then we give more details about the entire framework.

We first conduct LR feature extraction implemented by one $3 \times 3$ convolution with $64$ output channels. Then, the key component of our network utilizes multiple stacked information multi-distillation blocks (IMDBs) and assembles all intermediate features to fusing by a $1 \times 1$ convolution layer. This scheme,  intermediate information collection (IIC), is beneficial to guarantee the integrity of the collected information and can further boost the SR performance by increasing very few parameters. The final upsampler only consists of one learnable layer and a non-parametric operation (sub-pixel convolution) for saving parameters as much as possible.

\begin{figure}[htpb]
	\centering
	\includegraphics[width=0.25\textwidth, height=0.32\textheight]{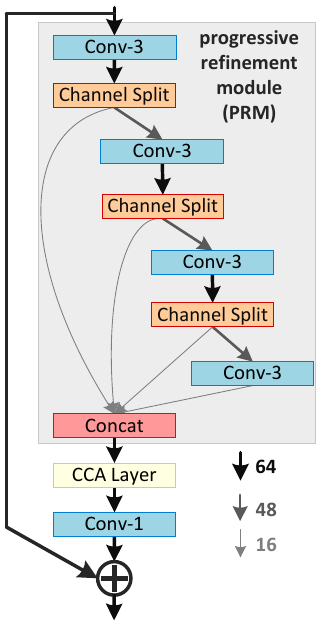}
	\vspace{-3mm}
	\caption{The architecture of our proposed information multi-distillation block (IMDB). Here, $64$, $48$, and $16$ all represent the output channels of the convolution layer. ``Conv-3'' denotes the $3 \times 3$ convolutional layer, and ``CCA Layer'' indicates the proposed contrast-aware channel attention (CCA) that is depicted in Figure~\ref{fig:cca}. Each convolution followed by a Leaky ReLU activation function except for the last $1 \times 1$ convolution. We omit them for concise.}
	\label{fig:IMDB}
\end{figure}

\subsection{Information multi-distillation block}
As depicted in Figure~\ref{fig:IMDB}, our information multi-distillation block (IMDB) is constructed by progressive refinement module, contrast-aware channel attention (CCA) layer, and a $1 \times 1$ convolution that is used to reduce the number of feature channels. The whole block adopts residual connection. The main idea of this block is extracting useful features little by little like DenseNet~\cite{DenseNet}. Then we give more details to these modules.
\subsubsection{Progressive refinement module}
\begin{table}[htpb]
	\centering
	\small
	\caption{PRM architecture. The columns represent layer, kernel-size, stride, input channels, and output channels. The symbols, C, and L denote a convolution layer, and Leaky ReLU ($\alpha  = 0.05$).}
	\begin{tabular}{ccccc}
		\hline
		Layer & Kernel & Stride & Input\_channel & Output\_channel \\
		\hline
		CL & 3 & 1 & 64 & 64 \\
		CL & 3 & 1 & 48 & 64 \\
		CL & 3 & 1 & 48 & 64 \\
		CL & 3 & 1 & 48 & 16 \\
		\hline
	\end{tabular}
	\label{tab:prm}
\end{table}

As labeled with the \textcolor{gray}{gray} box in Figure~\ref{fig:IMDB}, the progressive refinement module (PRM) first adopts the $3 \times 3$ convolution layer to extract input features for multiple subsequent distillation (refinement) steps. For each step, we employ channel split operation on the preceding features, which will produce two-part features. One is preserved and the other portion is fed into the next calculation unit. The retained part can be regarded as the refined features. Given the input features $F_{in}$, this procedure in the $n$-th IMDB can be described as
\begin{equation}
\begin{aligned}
&F_{refined\_1}^n,F_{coarse\_1}^n = Split_1^n\left( {CL_1^n\left( {F_{in}^n} \right)} \right), \\
&F_{refined\_2}^n,F_{coarse\_2}^n = Split_2^n\left( {CL_2^n\left( {F_{coarse\_1}^n} \right)} \right), \\
&F_{refined\_3}^n,F_{coarse\_3}^n = Split_3^n\left( {CL_3^n\left( {F_{coarse\_2}^n} \right)} \right), \\
&F_{refined\_4}^n = CL_4^n\left( {F_{coarse\_3}^n} \right),
\end{aligned} 
\end{equation}
where $CL_j^n$ denotes the $j$-th convolution layer (including Leaky ReLU) of the $n$-th IMDB, $Split_j^n$ indicates the $j$-th channel split layer of the $n$-th IMDB, $F_{refined\_j}^n$ represents the $j$-th refined features (preserved), and $F_{coarse\_j}^n$ is the $j$-th coarse features to be further processed. The hyperparameter of PRM architecture is shown in Table~\ref{tab:prm}. The following stage is concatenating refined features from each step. It can be expressed by
\begin{equation}
\begin{aligned}
F_{distilled}^n &= \\ Concat&\left( {F_{refined\_1}^n,F_{refined\_2}^n,F_{refined\_3}^n,F_{refined\_4}^n} \right),
\end{aligned}
\end{equation}
where $Concat$ denotes concatenation operation along the channel dimension.

\subsubsection{Contrast-aware channel attention layer}
\begin{figure}[htpb]
	\centering
	\includegraphics[width=0.24\textwidth]{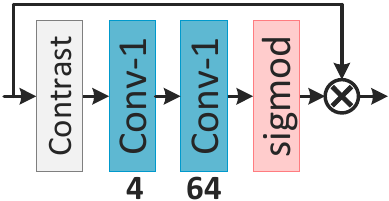}
	\vspace{-3mm}
	\caption{Contrast-aware channel attention module.}
	\label{fig:cca}
\end{figure}
The initial channel attention is employed in image classification task and is well-known as the squeeze-and-excitation (SE) module. In the high-level field, the importance of a feature map depends on activated high-value areas, since these regions in favor of classification or detection. Accordingly, global average/maximum pooling is utilized to capture the global information in these high-level or mid-level vision. Although the average pooling can indeed improve the PSNR value, it lacks the information about structures, textures, and edges that are propitious to enhance image details (related to SSIM). As depicted in Figure~\ref{fig:cca}, the contrast-aware channel attention module is special to low-level vision,~\eg, image super-resolution, and enhancement. Specifically, we replace global average pooling with the summation of standard deviation and mean (evaluating the contrast degree of a feature map). Let's denote $X = \left[ {{x_1}, \ldots ,{x_c}, \ldots ,{x_C}} \right]$ as the input, which has $C$ feature maps with spatial size of $H \times W$. Therefore, the contrast information value can be calculated by
\begin{equation}
\begin{aligned}
{z_c} &= {H_{GC}}\left( {{x_c}} \right) \\&= \sqrt {\frac{1}{{HW}} {\sum\limits_{\left( {i,j} \right) \in {x_c}} {{{\left( {x_c^{i,j} - \frac{1}{{HW}}\sum\limits_{\left( {i,j} \right) \in {x_c}} {x_c^{i,j}} } \right)}^2}} } } + \\ 
&\frac{1}{{HW}}\sum\limits_{\left( {i,j} \right) \in {x_c}} {x_c^{i,j}},
\end{aligned}
\end{equation}
where ${z_c}$ is the $c$-th element of output. ${H_{GC}}\left(  \cdot  \right)$ indicates the global contrast (GC) information evaluation function. With the assistance of the CCA module, our network can steadily improve the accuracy of SISR.

\subsection{Adaptive cropping strategy}

\begin{figure}[ht]
	\centering
	\subfigure[The first image patch]{\includegraphics[width=0.2\textwidth]{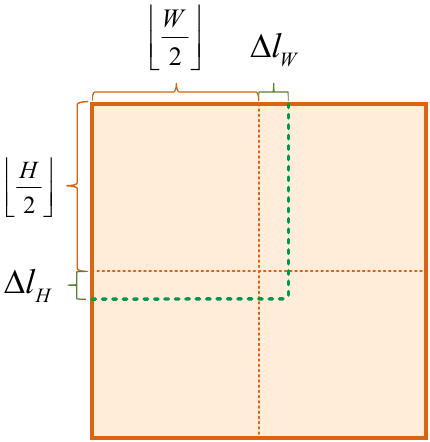}}
	\hfil
	\subfigure[The last image patch]{\includegraphics[width=0.23\textwidth]{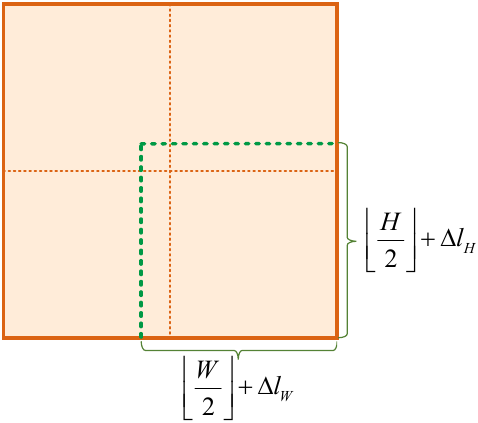}}
	\vspace{-3mm}
	\caption{The diagrammatic sketch of adaptive cropping strategy (ACS). The cropped image patches in the green dotted boxes.}
	\label{fig:acs}
\end{figure}

\begin{figure}[htpb]
	\centering
	\includegraphics[width=0.45\textwidth]{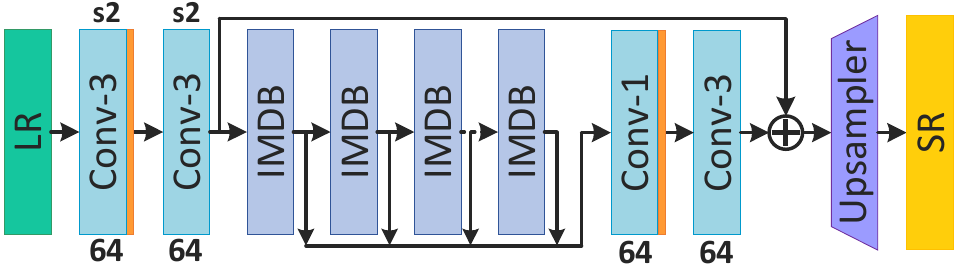}
	\vspace{-3mm}
	\caption{The network structure of our IMDN\_AS. ``s2'' represents the stride of 2.}
	\label{fig:IMDN-anyscale}
\end{figure}

The adaptive cropping strategy (ACS) is special to image of any arbitrary size super-resolving. Meanwhile, it can also deal with the SR problem of any scale factor with a single model (see Figure~\ref{fig:IMDN-anyscale}). We slightly modify the original IMDN by introducing two downsampling layer and construct the current IMDN\_AS (IMDN for any scales). Here, the LR and HR images have the same spatial size (height and width). To handle images whose height and width are not divisible by $4$, we first cut the entire images into $4$ parts and then feed them into our IMDN\_AS. As illustrated in Figure~\ref{fig:acs}, we can obtain $4$ overlapped image patches through ACS. Take the first patch in the upper left corner as an example, and we give the details about ACS. This image patch must satisfy
\begin{equation}
\begin{aligned}
&{\left( {\left\lfloor {\frac{H}{2}} \right\rfloor  + \Delta {l_H}} \right)\% 4 = 0}, \\
&{\left( {\left\lfloor {\frac{W}{2}} \right\rfloor  + \Delta {l_W}} \right)\% 4 = 0},
\end{aligned}
\end{equation}
where $\Delta {l_H}$, $\Delta {l_W}$ are extra increments of height and width, respectively. They can be computed by
\begin{equation}
\begin{aligned}
&\Delta {l_H} = {padding}_H - \left( {\left\lfloor {\frac{H}{2}} \right\rfloor  + {padding}_H} \right)\% 4, \\
&\Delta {l_W} = {padding}_W - \left( {\left\lfloor {\frac{W}{2}} \right\rfloor  + {padding}_W} \right)\% 4,
\end{aligned}
\end{equation}
where ${padding}_H$, ${padding}_W$ are preset additional lengths. In general, their values are setting by
\begin{equation}
{padding}_H = {padding}_W = 4k, k \ge 1.
\end{equation}
Here, $k$ is an integer greater than or equal to 1. These four patches can be processed in parallel (they have the same sizes), after which the outputs are pasted to their original location, and the extra increments ($\Delta {l_H}$ and $\Delta {l_W}$) are discarded.

\section{Experiments}\label{sec:experiemnts}
\subsection{Datasets and metrics}
In our experiments, we use the DIV2K dataset~\cite{NTIRE2017_dataset}, which contains 800 high-quality RGB training images and widely used in image restoration tasks~\cite{EDSR,RDN,RCAN,RNAN}. For evaluation, we use five widely used benchmark datasets: Set5~\cite{Set5}, Set14~\cite{Set14}, BSD100~\cite{BSD100}, Urban100~\cite{Urban100}, and Manga109~\cite{Manga109}. We evaluate the performance of the super-resolved images using two metrics, including peak signal-to-noise ratio (PSNR) and structure similarity index (SSIM)~\cite{SSIM}. As with existing works~\cite{VDSR,DRRN,EDSR,RDN,RCAN,IDN,CARN}, we calculate the values on the luminance channel (\ie, Y channel of the YCbCr channels converted from the RGB channels).

Additionally, for any/unknown scale factor experiments, we use RealSR dataset from NTIRE2019 Real Super-Resolution Challenge\footnote{\url{http://www.vision.ee.ethz.ch/ntire19/}}. It is a novel dataset of real low and high resolution paired images. The training data consists of 60 real low, and high resolution paired images, and the validation data contains 20 LR-HR pairs. It is noteworthy that the LR and HR have the same size. 
\subsection{Implementation details}
To obtain LR DIV2K training images, we downscale HR images with the scaling factors ($\times 2$, $\times 3$, and $\times 4$) using bicubic interpolation in MATLAB R2017a. The HR image patches with a size of $192 \times 192$ are randomly cropped from HR images as the input of our model, and the mini-batch size is set to $16$. For data augmentation, we perform randomly horizontal flip and $90$ degree rotation. Our model is trained by ADAM optimizer with the momentum parameter ${\beta _1} = 0.9$. The initial learning rate is set to $2 \times {10^{ - 4}}$ and halved at every $2 \times {10^5}$ iterations. We set the number of IMDB to $6$ in our IMDN and IMDN\_AS. We apply PyTorch framework to implement the proposed network on the desktop computer with 4.2GHz Intel i7-7700K CPU, 64G RAM, and NVIDIA TITAN Xp GPU (12G memory).

\subsection{Model analysis}
In this subsection, we investigate model parameters, the effectiveness of IMDB, the intermediate information collection scheme, and adaptive cropping strategy.
\subsubsection{Model parameters}
\begin{figure}[htpb]
	\centering
	\includegraphics[width=0.4\textwidth]{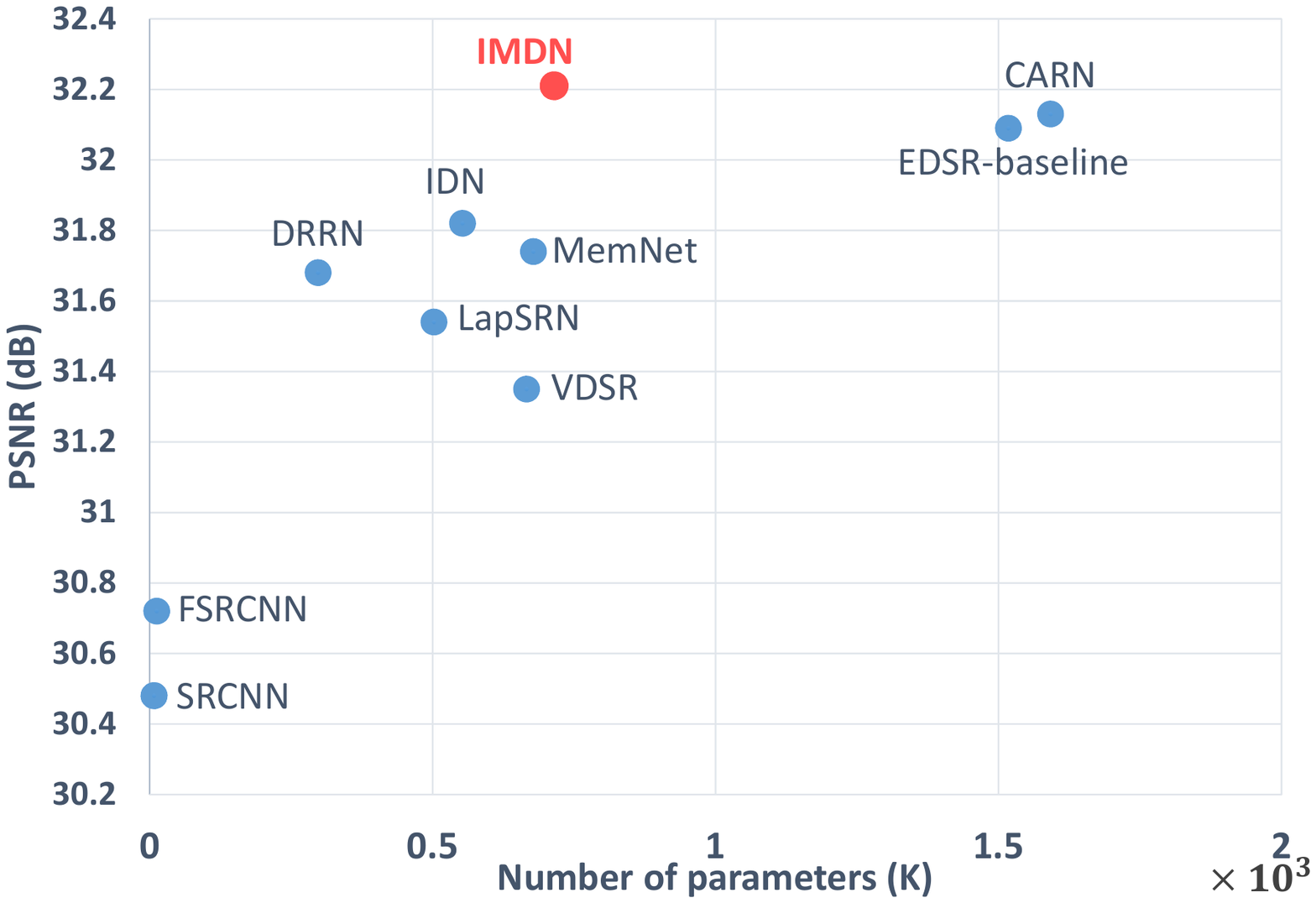}
	\vspace{-3mm}
	\caption{Trade-off between performance and number of parameters on Set5 $\times 4$ dataset.}
	\label{fig:parameters}
\end{figure}
To construct a lightweight SR model, the parameters of the network is vital. From Table~\ref{tab:psnr-ssim}, we can observe that our IMDN with fewer parameters achieves comparative or better performance when comparing with other state-of-the-art methods, such as EDSR-baseline (CVPRW'17), IDN (CVPR'18), SRMDNF (CVPR'18), and CARN (ECCV'18). We also visualize the trade-off analysis between performance and model size in Figure~\ref{fig:parameters}. We can see that our IMDN achieves a better trade-off between the performance and model size.

\subsubsection{Ablation studies of CCA module and IIC scheme}
\begin{table*}[htpb]
	\centering
	\caption{Investigations of CCA module and IIC scheme.}
	\label{tab:cca-iic}
	\begin{tabular}{|c|c|c|c|c|c|c|c|c|c|}
		\hline
		\multirow{2}{*}{Scale} & \multirow{2}{*}{PRM} & \multirow{2}{*}{CCA} & \multirow{2}{*}{IIC} & \multirow{2}{*}{Params} & Set5 & Set14 & BSD100 & Urban100 & Manga109 \\
		\cline{6-10}
		& & & & & PSNR / SSIM & PSNR / SSIM & PSNR / SSIM & PSNR / SSIM & PSNR / SSIM \\
		\hline
		\hline
		\multirow{4}{*}{$\times 4$} & \XSolid & \XSolid & \XSolid & 510K & 31.86 / 0.8901 & 28.43 / 0.7775 &  27.45 / 0.7320 & 25.63 / 0.7711 & 29.92 / 0.9003 \\
		
		& \Checkmark & \XSolid & \XSolid & 480K & 32.01 / 0.8927 & 28.49 / 0.7792 & 27.50 / 0.7338 & 25.81 / 0.7773 & 30.16 / 0.9038 \\
		& \Checkmark& \Checkmark & \XSolid & 482K & 32.10 / 0.8934 & 28.51 / 0.7794 & 27.52 / 0.7341 & 25.89 / 0.7793 & 30.25 / 0.9050 \\
		& \Checkmark & \Checkmark & \Checkmark & 499K & \textbf{32.11} / \textbf{0.8934} & \textbf{28.52} / \textbf{0.7797} & \textbf{27.53} / \textbf{0.7342} & \textbf{25.90} / \textbf{0.7797} & \textbf{30.28} / \textbf{0.9054} \\
		\hline
	\end{tabular}
\end{table*}

\begin{table}[htpb]
	\small
	\centering
	\caption{Comparison with original channel attention (CA) and the presented contrast-aware channel attention (CCA).}
	\begin{tabular}{|l|c|c|c|c|}
		\hline
		Module & Set5 & Set14 & BSD100 & Urban100\\
		\hline
		\hline
		IMDN\_basic\_B4 + CA & 32.0821 & 28.5086 & 27.5124 & 25.8829 \\
		IMDN\_basic\_B4 + CCA & \textbf{32.0964} & \textbf{28.5118} & \textbf{27.5185} & \textbf{25.8916} \\
		\hline
	\end{tabular}
	\label{tab:ca-and-cca}
\end{table}

\begin{figure}
	\centering
	\includegraphics[width=0.38\textwidth]{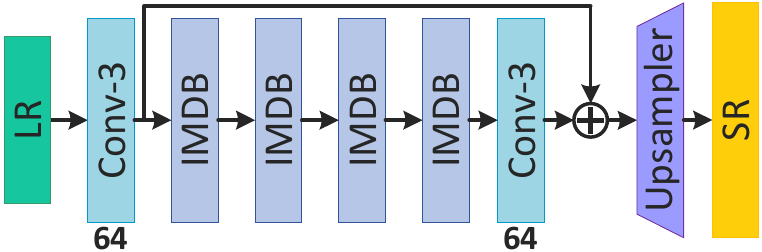}
	\caption{The structure of IMDN\_basic\_B4.}
	\label{fig:IMDN-basic-B4}
	\vspace{-3mm}
\end{figure}

To quickly validate the effectiveness of the contrast-aware attention (CCA) module and intermediate information collection (IIC) scheme, we adopt $4$ IMDBs to conduct the following ablation study experiment, named IMDN\_B4. When removing the CCA module and IIC scheme, the IMDN\_B4 becomes IMDN\_basic\_B4 as illustrated in Figure~\ref{fig:IMDN-basic-B4}. From Table~\ref{tab:cca-iic}, we can find out that the CCA module leads to performance improvement (PSNR: \textbf{+0.09dB}, SSIM: \textbf{+0.0012} for $\times 4$ Manga109) only by increasing 2K parameters (which is an increase of $0.4\% $). The results compared with the CA module are placed in Table~\ref{tab:ca-and-cca}. To study the efficiency of PRM in IMDB, we replace it with three cascaded $3 \times 3$ convolution layers (64 channels) and remove the final $1 \times 1$ convolution (used for fusion). The compared results are given in Table~\ref{tab:cca-iic}. Although this network has more parameters (510K), its performance is much lower than our IMDN\_basic\_B4 (480K) especially on Urban100 and Manga109 datasets.

\subsubsection{Investigation of ACS}
\begin{table}[htpb]
	\centering
	\small
	\caption{Quantitative evaluation of VDSR and our IMDN\_AS in PSNR, SSIM, LPIPS, running time, and memory occupation.}
	\label{fig:study-acs}
	\begin{tabular}{|c|c|c|c|c|c|}
		\hline
		Method & PSNR & SSIM & LPIPS~\cite{LPIPS} & Time & Memory \\
		\hline
		VDSR~\cite{VDSR} & 28.75 & 0.8439 & 0.2417 & 0.0290 & 7,855M \\ 
		IMDN\_AS & \textbf{29.35} & \textbf{0.8595} & \textbf{0.2147} & \textbf{0.0041} & \textbf{3,597M} \\
		\hline
	\end{tabular}
\end{table}

To verify the efficiency of the proposed adaptive cropping strategy (ACS), we use RealSR training images to train VDSR~\cite{VDSR} and our IMDN\_AS. The results, evaluated on RealSR RGB validation dataset, are illustrated in Table~\ref{fig:study-acs} and we can easily observe that the presented IMDN\_AS achieves better performance in term of image quality, execution speed, and footprint. Accordingly, it also suggests the proposed ACS is powerful to address SR problem of any scales.

\subsection{Comparison with state-of-the-arts}
We compare our IMDN with 11 state-of-the-art methods: SRCNN~\cite{SRCNN,SRCNN-Ex}, FSRCNN~\cite{FSRCNN}, VDSR~\cite{VDSR}, DRCN~\cite{DRCN}, LapSRN~\cite{LapSRN}, DRRN~\cite{DRRN}, MemNet~\cite{MemNet}, IDN~\cite{IDN}, EDSR-baseline~\cite{EDSR}, SRMDNF~\cite{SRMDNF}, and CARN~\cite{CARN}. Table~\ref{tab:psnr-ssim} shows quantitative comparisons for $\times 2$, $\times 3$, and $\times4$ SR. It can find out that our IMDN performs favorably against other compared approaches on most datasets, especially at the scaling factor of $\times 2$. 

\begin{figure*}[htpb]
	\scriptsize
	\centering
	\scalebox{0.84}{
		\begin{tabular}{lc}
			\begin{adjustbox}{valign=t}
				\begin{tabular}{c}
					\includegraphics[width=0.36\textwidth, height=0.226\textheight]{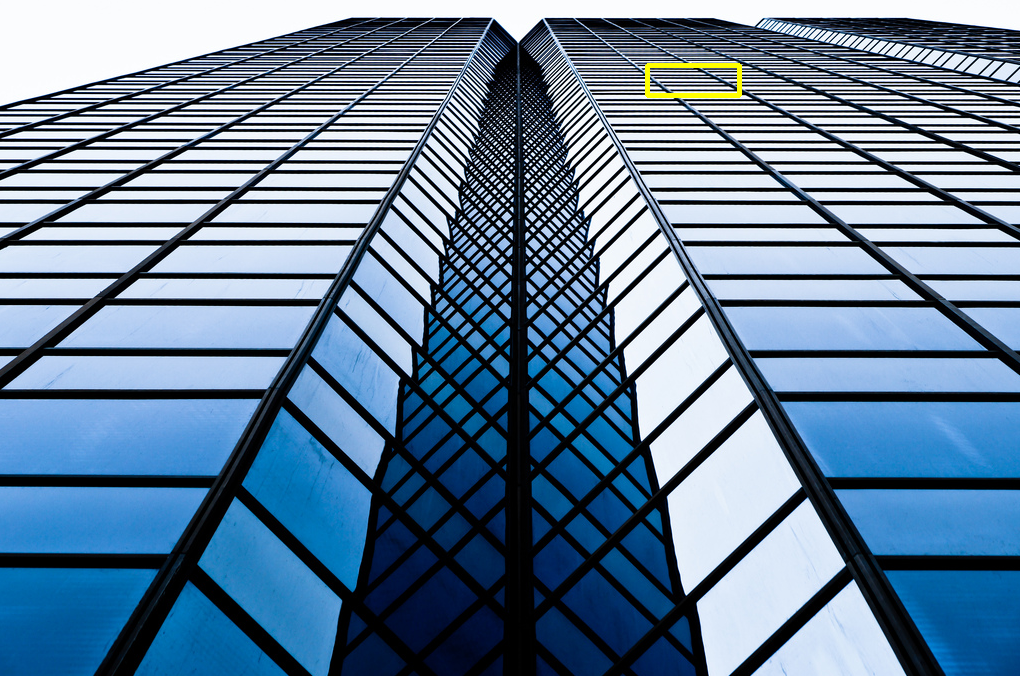} \\
					Urban100 ($2 \times$): \\
					img\_67 \\
				\end{tabular}
			\end{adjustbox}
			\hspace{-3mm}
			\begin{adjustbox}{valign=t}
				\begin{tabular}{ccccc}
					\includegraphics[width=0.14\textwidth, height=0.1\textheight]{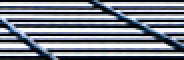} & 
					\hspace{-3mm}
					\includegraphics[width=0.14\textwidth, height=0.1\textheight]{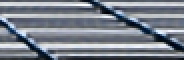} & 
					\hspace{-3mm}
					\includegraphics[width=0.14\textwidth, height=0.1\textheight]{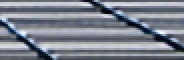} & 
					\hspace{-3mm}
					\includegraphics[width=0.14\textwidth, height=0.1\textheight]{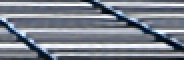} & 
					\hspace{-3mm}
					\includegraphics[width=0.14\textwidth, height=0.1\textheight]{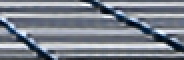} \\
					HR & \hspace{-3mm}
					VDSR~\cite{VDSR} & \hspace{-3mm}
					DRCN~\cite{DRCN} & \hspace{-3mm}
					DRRN~\cite{DRRN} & \hspace{-3mm}
					LapSRN~\cite{LapSRN} \\
					PSNR/SSIM & \hspace{-3mm}
					24.10/0.9537 & \hspace{-3mm}
					23.64/0.9493 & \hspace{-3mm}
					24.73/0.9594 & \hspace{-3mm}
					23.80/0.9527 \\
					\includegraphics[width=0.14\textwidth, height=0.1\textheight]{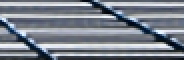} & 
					\hspace{-3mm}
					\includegraphics[width=0.14\textwidth, height=0.1\textheight]{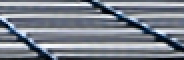} &
					\hspace{-3mm}
					\includegraphics[width=0.14\textwidth, height=0.1\textheight]{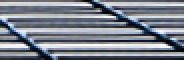} &
					\hspace{-3mm}
					\includegraphics[width=0.14\textwidth, height=0.1\textheight]{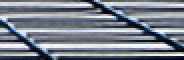} &
					\hspace{-3mm}
					\includegraphics[width=0.14\textwidth, height=0.1\textheight]{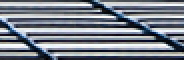} \\
					MemNet~\cite{MemNet} & \hspace{-3mm}
					IDN~\cite{IDN} & \hspace{-3mm}
					EDSR-baseline~\cite{EDSR} & \hspace{-3mm}
					CARN~\cite{CARN} & \hspace{-3mm}
					IMDN (Ours) \\
					24.98/0.9613 & \hspace{-3mm}
					24.68/0.9594 & \hspace{-3mm}
					26.01/0.9695 & \hspace{-3mm}
					25.96/0.9692 & \hspace{-3mm}
					\textbf{27.75}/\textbf{0.9773} \\
				\end{tabular}
			\end{adjustbox}
			\\
			
			\begin{adjustbox}{valign=t}
				\begin{tabular}{c}
					\includegraphics[width=0.36\textwidth, height=0.226\textheight]{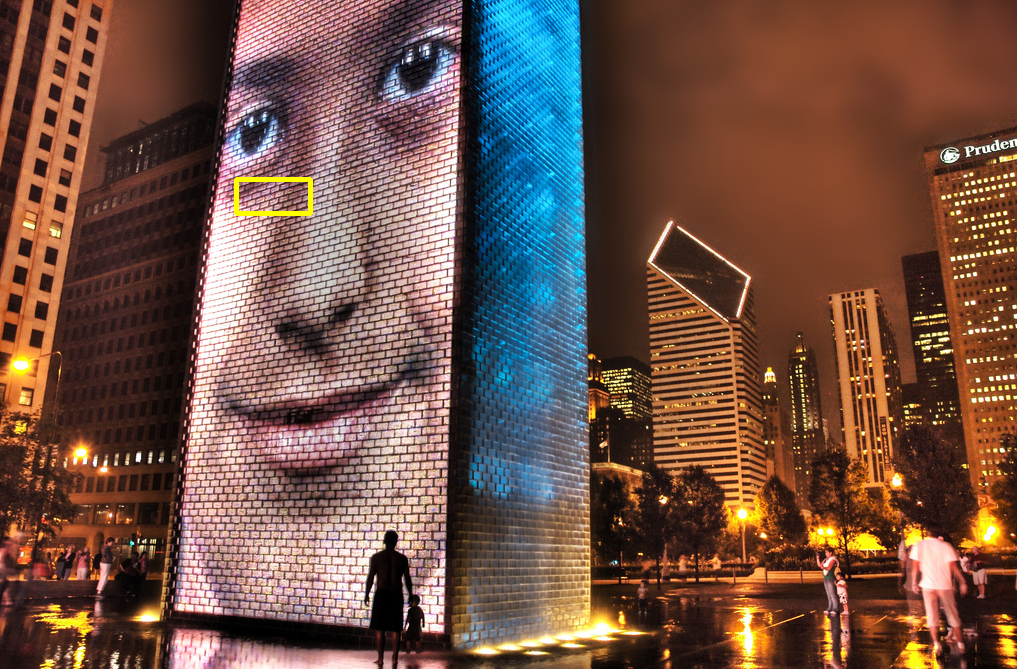} \\
					Urban100 ($3 \times$): \\
					img\_76 \\
				\end{tabular}
			\end{adjustbox}
			\hspace{-3mm}
			\begin{adjustbox}{valign=t}
				\begin{tabular}{ccccc}
					\includegraphics[width=0.14\textwidth, height=0.1\textheight]{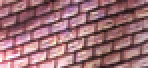} & 
					\hspace{-3mm}
					\includegraphics[width=0.14\textwidth, height=0.1\textheight]{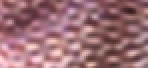} & 
					\hspace{-3mm}
					\includegraphics[width=0.14\textwidth, height=0.1\textheight]{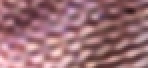} & 
					\hspace{-3mm}
					\includegraphics[width=0.14\textwidth, height=0.1\textheight]{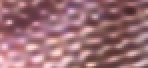} & 
					\hspace{-3mm}
					\includegraphics[width=0.14\textwidth, height=0.1\textheight]{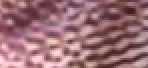} \\
					HR & \hspace{-3mm}
					VDSR~\cite{VDSR} & \hspace{-3mm}
					DRCN~\cite{DRCN} & \hspace{-3mm}
					DRRN~\cite{DRRN} & \hspace{-3mm}
					LapSRN~\cite{LapSRN} \\
					PSNR/SSIM & \hspace{-3mm}
					24.75/0.8284 & \hspace{-3mm}
					24.82/0.8277 & \hspace{-3mm}
					24.80/0.8312 & \hspace{-3mm}
					24.89/0.8337 \\
					\includegraphics[width=0.14\textwidth, height=0.1\textheight]{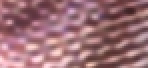} & 
					\hspace{-3mm}
					\includegraphics[width=0.14\textwidth, height=0.1\textheight]{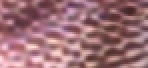} &
					\hspace{-3mm}
					\includegraphics[width=0.14\textwidth, height=0.1\textheight]{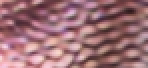} &
					\hspace{-3mm}
					\includegraphics[width=0.14\textwidth, height=0.1\textheight]{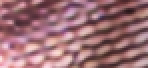} &
					\hspace{-3mm}
					\includegraphics[width=0.14\textwidth, height=0.1\textheight]{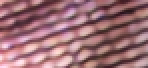} \\
					MemNet~\cite{MemNet} & \hspace{-3mm}
					IDN~\cite{IDN} & \hspace{-3mm}
					EDSR-baseline~\cite{EDSR} & \hspace{-3mm}
					CARN~\cite{CARN} & \hspace{-3mm}
					IMDN (Ours) \\
					24.97/0.8359 & \hspace{-3mm}
					24.95/0.8332 & \hspace{-3mm}
					25.85/0.8565 & \hspace{-3mm}
					25.92/0.8583 & \hspace{-3mm}
					\textbf{26.19}/\textbf{0.8610} \\
				\end{tabular}
			\end{adjustbox}
			\\
			
	\end{tabular} }
	\vspace{-3mm}
	\caption{Visual comparisons of IMDN with other SR methods on Set5 and Urban100 datasets.}
	\label{fig:visual-comparison}
\end{figure*}

Figure~\ref{fig:visual-comparison} shows $\times 2$, $\times 3$ and $\times 4$ visual comparisons on Set5 and Urban100 datasets. For ``img\_67'' image from Urban100, we can see that grid structure is recovered better than others. It also demonstrates the effectiveness of our IMDN.
\begin{table*}[htpb]
	\caption{Average PSNR/SSIM for scale factor $\times 2$, $\times 3$ and $\times 4$ on datasets Set5, Set14, BSD100, Urban100, and Manga109. Best and second best results are \textbf{highlighted} and \underline{underlined}.}
	\centering
	\begin{tabular}{|l|c|c|c|c|c|c|c|}
		\hline
		\multirow{2}{*}{Method} & \multirow{2}{*}{Scale} & \multirow{2}{*}{Params} & Set5 & Set14 & BSD100 & Urban100 & Manga109 \\
		\cline{4-8}
		& & & PSNR / SSIM & PSNR / SSIM & PSNR / SSIM & PSNR / SSIM & PSNR / SSIM \\
		\hline
		\hline
		Bicubic & \multirow{13}{*}{$\times 2$} & - &33.66 / 0.9299 & 30.24 / 0.8688 & 29.56 / 0.8431 & 26.88 / 0.8403 & 30.80 / 0.9339 \\
		
		SRCNN~\cite{SRCNN} & & 8K & 36.66 / 0.9542 & 32.45 / 0.9067 & 31.36 / 0.8879 & 29.50 / 0.8946 & 35.60 / 0.9663 \\
		
		FSRCNN~\cite{FSRCNN} & & 13K & 37.00 / 0.9558 & 32.63 / 0.9088 & 31.53 / 0.8920 & 29.88 / 0.9020 & 36.67 / 0.9710 \\
		
		VDSR~\cite{VDSR} & & 666K & 37.53 / 0.9587 & 33.03 / 0.9124 & 31.90 / 0.8960 & 30.76 / 0.9140 & 37.22 / 0.9750 \\
		
		DRCN~\cite{DRCN} &  & 1,774K & 37.63 / 0.9588 & 33.04 / 0.9118 & 31.85 / 0.8942 & 30.75 / 0.9133 & 37.55 / 0.9732 \\
		
		LapSRN~\cite{LapSRN} &  & 251K & 37.52 / 0.9591 & 32.99 / 0.9124 & 31.80 / 0.8952 & 30.41 / 0.9103 & 37.27 / 0.9740 \\
		
		DRRN~\cite{DRRN} &  & 298K & 37.74 / 0.9591 & 33.23 / 0.9136 & 32.05 / 0.8973 & 31.23 / 0.9188 & 37.88 / 0.9749 \\
		
		MemNet~\cite{MemNet} &  & 678K & 37.78 / 0.9597 & 33.28 / 0.9142 & 32.08 / 0.8978 & 31.31 / 0.9195 & 37.72 / 0.9740 \\
		
		IDN~\cite{IDN} &  & 553K & 37.83 / 0.9600 & 33.30 / 0.9148 & 32.08 / 0.8985 & 31.27 / 0.9196 & 38.01 / 0.9749 \\
		
		EDSR-baseline~\cite{EDSR} &  & 1,370K &\underline{37.99} / \underline{0.9604} & \underline{33.57} / \underline{0.9175} & \underline{32.16} / \underline{0.8994} & \underline{31.98} / \underline{0.9272} & \underline{38.54} / \underline{0.9769} \\
		
		SRMDNF~\cite{SRMDNF} &  & 1,511K & 37.79 / 0.9601 & 33.32 / 0.9159 & 32.05 / 0.8985 & 31.33 / 0.9204 & 38.07 / 0.9761 \\
		
		CARN~\cite{CARN} &  & 1,592K & 37.76 / 0.9590 & 33.52 / 0.9166 & 32.09 / 0.8978 & 31.92 / 0.9256 & 38.36 / 0.9765 \\
		
		IMDN (Ours) &  & 694K & \textbf{38.00} / \textbf{0.9605} & \textbf{33.63} / \textbf{0.9177} & \textbf{32.19} / \textbf{0.8996} & \textbf{32.17} / \textbf{0.9283} & \textbf{38.88} / \textbf{0.9774} \\
		\hline
		\hline
		Bicubic & \multirow{13}{*}{$\times 3$} & - & 30.39 / 0.8682 & 27.55 / 0.7742 & 27.21 / 0.7385 & 24.46 / 0.7349 & 26.95 / 0.8556 \\
		
		SRCNN~\cite{SRCNN} &  & 8K & 32.75 / 0.9090 & 29.30 / 0.8215 & 28.41 / 0.7863 & 26.24 / 0.7989 & 30.48 / 0.9117\\
		
		FSRCNN~\cite{FSRCNN} &  & 13K & 33.18 / 0.9140 & 29.37 / 0.8240 & 28.53 / 0.7910 & 26.43 / 0.8080 & 31.10 / 0.9210 \\
		
		VDSR~\cite{VDSR} &  & 666K & 33.66 / 0.9213 & 29.77 / 0.8314 & 28.82 / 0.7976 & 27.14 / 0.8279 & 32.01 / 0.9340 \\
		
		DRCN~\cite{DRCN} &  & 1,774K & 33.82 / 0.9226 & 29.76 / 0.8311 & 28.80 / 0.7963 & 27.15 / 0.8276 & 32.24 / 0.9343 \\
		
		LapSRN~\cite{LapSRN} &  & 502K & 33.81 / 0.9220 & 29.79 / 0.8325 & 28.82 / 0.7980 & 27.07 / 0.8275 & 32.21 / 0.9350 \\
		
		DRRN~\cite{DRRN} &  & 298K & 34.03 / 0.9244 & 29.96 / 0.8349 & 28.95 / 0.8004 & 27.53 / 0.8378 & 32.71 / 0.9379 \\
		
		MemNet~\cite{MemNet} &  & 678K & 34.09 / 0.9248 & 30.00 / 0.8350 & 28.96 / 0.8001 & 27.56 / 0.8376 & 32.51 / 0.9369\\
		
		IDN~\cite{IDN} &  & 553K & 34.11 / 0.9253 & 29.99 / 0.8354 & 28.95 / 0.8013 & 27.42 / 0.8359 & 32.71 / 0.9381 \\
		
		EDSR-baseline~\cite{EDSR} &  & 1,555K & \textbf{34.37} / \textbf{0.9270} & 30.28 / \textbf{0.8417} & \textbf{29.09} / \textbf{0.8052} & \underline{28.15} / \textbf{0.8527} & 33.45 / 0.9439 \\
		
		SRMDNF~\cite{SRMDNF} &  & 1,528K & 34.12 / 0.9254 & 30.04 / 0.8382 & 28.97 / 0.8025 & 27.57 / 0.8398 & 33.00 / 0.9403 \\
		
		CARN~\cite{CARN} &  & 1,592K & 34.29 / 0.9255 & \underline{30.29} / 0.8407 & \underline{29.06} / 0.8034 & 28.06 / 0.8493 & \underline{33.50} / \underline{0.9440} \\
		
		IMDN (Ours) &  & 703K & \underline{34.36} / \textbf{0.9270} & \textbf{30.32} / \textbf{0.8417} & \textbf{29.09} / \underline{0.8046} & \textbf{28.17} / \underline{0.8519} & \textbf{33.61} / \textbf{0.9445} \\
		
		\hline
		\hline
		Bicubic & \multirow{13}{*}{$\times 4$} & - & 28.42 / 0.8104 & 26.00 / 0.7027 & 25.96 / 0.6675 & 23.14 / 0.6577 & 24.89 / 0.7866 \\
		
		SRCNN~\cite{SRCNN} &  & 8K & 30.48 / 0.8628 & 27.50 / 0.7513 & 26.90 / 0.7101 & 24.52 / 0.7221 & 27.58 / 0.8555 \\
		
		FSRCNN~\cite{FSRCNN} &  & 13K & 30.72 / 0.8660 & 27.61 / 0.7550 & 26.98 / 0.7150 & 24.62 / 0.7280 &  27.90 / 0.8610 \\
		
		VDSR~\cite{VDSR} &  & 666K & 31.35 / 0.8838 & 28.01 / 0.7674 & 27.29 / 0.7251 & 25.18 / 0.7524 & 28.83 / 0.8870 \\
		
		DRCN~\cite{DRCN} &  & 1,774K & 31.53 / 0.8854 & 28.02 / 0.7670 & 27.23 / 0.7233 & 25.14 / 0.7510 & 28.93 / 0.8854 \\
		
		LapSRN~\cite{LapSRN} &  & 502K & 31.54 / 0.8852 & 28.09 / 0.7700 & 27.32 / 0.7275 & 25.21 / 0.7562 & 29.09 / 0.8900 \\
		
		DRRN~\cite{DRRN} &  & 298K & 31.68 / 0.8888 & 28.21 / 0.7720 & 27.38 / 0.7284 & 25.44 / 0.7638 & 29.45 / 0.8946 \\
		
		MemNet~\cite{MemNet} &  & 678K & 31.74 / 0.8893 & 28.26 / 0.7723 & 27.40 / 0.7281 & 25.50 / 0.7630 & 29.42 / 0.8942 \\
		
		IDN~\cite{IDN} &  & 553K & 31.82 / 0.8903 & 28.25 / 0.7730 & 27.41 / 0.7297 &  25.41 / 0.7632 & 29.41 / 0.8942 \\
		
		EDSR-baseline~\cite{EDSR} &  & 1,518K & 32.09 / \underline{0.8938} &  \underline{28.58} / \textbf{0.7813} & \underline{27.57} / \textbf{0.7357} & 26.04 /  \textbf{0.7849} & 30.35 / 0.9067 \\
		
		SRMDNF~\cite{SRMDNF} &  & 1,552K & 31.96 / 0.8925 & 28.35 / 0.7787 & 27.49 / 0.7337 & 25.68 / 0.7731 & 30.09 / 0.9024\\
		
		CARN~\cite{CARN} &  & 1,592K & \underline{32.13} / 0.8937 & \textbf{28.60} / 0.7806 &  \textbf{27.58} / 0.7349 & \textbf{26.07} / 0.7837  & \textbf{30.47} / \textbf{0.9084} \\
		
		IMDN (Ours) &  & 715K & \textbf{32.21} / \textbf{0.8948} & \underline{28.58} / \underline{0.7811} & 27.56 / \underline{0.7353} & \underline{26.04} / \underline{0.7838} & \underline{30.45} / \underline{0.9075} \\
		\hline
		
	\end{tabular}
	\label{tab:psnr-ssim}
\end{table*}

\begin{table*}[htpb]
	\caption{Memory Consumption (MB) and average inference time (second).}
	\label{tab:memory-time}
	\begin{center}
		\begin{tabular}{|l|c|c|c|r|r|r|}
			\hline
			\multirow{2}{*}{Method} & \multirow{2}{*}{Scale} & \multirow{2}{*}{Params} & \multirow{2}{*}{Depth} & BSD100 & Urban100 & Manga109 \\
			\cline{5-7}
			& & & & Memory / Time & Memory / Time & Memory / Time \\
			\hline
			\hline
			EDSR-baseline~\cite{EDSR} & \multirow{6}{*}{$\times 4$} & 1.6M & 37 & 665 / 0.00295 & 2,511 / 0.00242 & 1,219 / 0.00232 \\
			
			EDSR~\cite{EDSR} &  & 43M & 69 & 1,531 / 0.00580 & 8,863 / 0.00416 & 3,703 / 0.00380 \\
			
			RDN~\cite{RDN} &  & 22M & 150 & 1,123 / 0.01626 & 3,335 / 0.01325 & 2,257 / 0.01300 \\
			
			RCAN~\cite{RCAN} &  & 16M & 415 & 777 / 0.09174 & 2,631 / 0.55280 & 1,343 / 0.72250 \\
			
			CARN~\cite{CARN} &  & 1.6M & 34 & 945 / 0.00278 & 3,761 / 0.00305 & 2,803 / 0.00383 \\
			
			IMDN (Ours) & & 0.7M & 34 & 671 / 0.00285 & 1,155 / 0.00284 & 895 / 0.00279 \\
			
			\hline
		\end{tabular}
	\end{center}
\end{table*}

\subsection{Running time}

\begin{table}[htpb]
	\centering
	\small
	\vspace{-3mm}
	\caption{The computational costs. For representing concisely, we omit $m^{2}$. Least and second least computational costs are \textbf{highlighted} and \underline{underlined}.}
	\label{tab:computational-costs}
	\begin{tabular}{|c|c|c|c|c|c|}
		\hline
		Scale & LapSRN~\cite{LapSRN} & IDN~\cite{IDN} & EDSR-b~\cite{EDSR} & CARN~\cite{CARN} & IMDN \\
		\hline
		\hline
		$\times 2$ & \textbf{112K} & 175K & 341K & \underline{157K} & 173K \\
		$\times 3$ & \underline{76K} & \textbf{75K} & 172K & 90K & 78K \\
		$\times 4$ & 76K & \underline{51K} & 122K & 76K & \textbf{45K} \\
		\hline
	\end{tabular}
\end{table}

\subsubsection{Complexity analysis}
As the proposed IMDN mainly consists of convolutions, the total number of parameters can be computed as
\begin{equation}
Params = \sum\limits_{l = 1}^L {\underbrace {{n_{l - 1}} \cdot {n_l} \cdot f_l^2}_{conv} + \underbrace {{n_l}}_{bias}} ,
\end{equation}
where $l$ is the layer index, $L$ denotes the total number of layers, and $f$ represents the spatial size of the filters. The number of convolutional kernels belong to $l$-th layer is $n_l$, and its input channels are $n_{l-1}$. Suppose that the spatial size of output feature maps is $m_l \times m_l$, the time complexity can be roughly calculated by
\begin{equation}\label{eq:time-complexity}
O\left( {\sum\limits_{l = 1}^L {{n_{l - 1}} \cdot {n_l} \cdot f_l^2 \cdot m_l^2} } \right).
\end{equation}
We assume that the size of the HR image is $m \times m$ and then the computational costs can be calculated by Equation~\ref{eq:time-complexity} (see Table~\ref{tab:computational-costs}).

\subsubsection{Running Time}
\begin{figure}[htpb]
	\centering
	\includegraphics[width=0.4\textwidth, height=0.21\textheight]{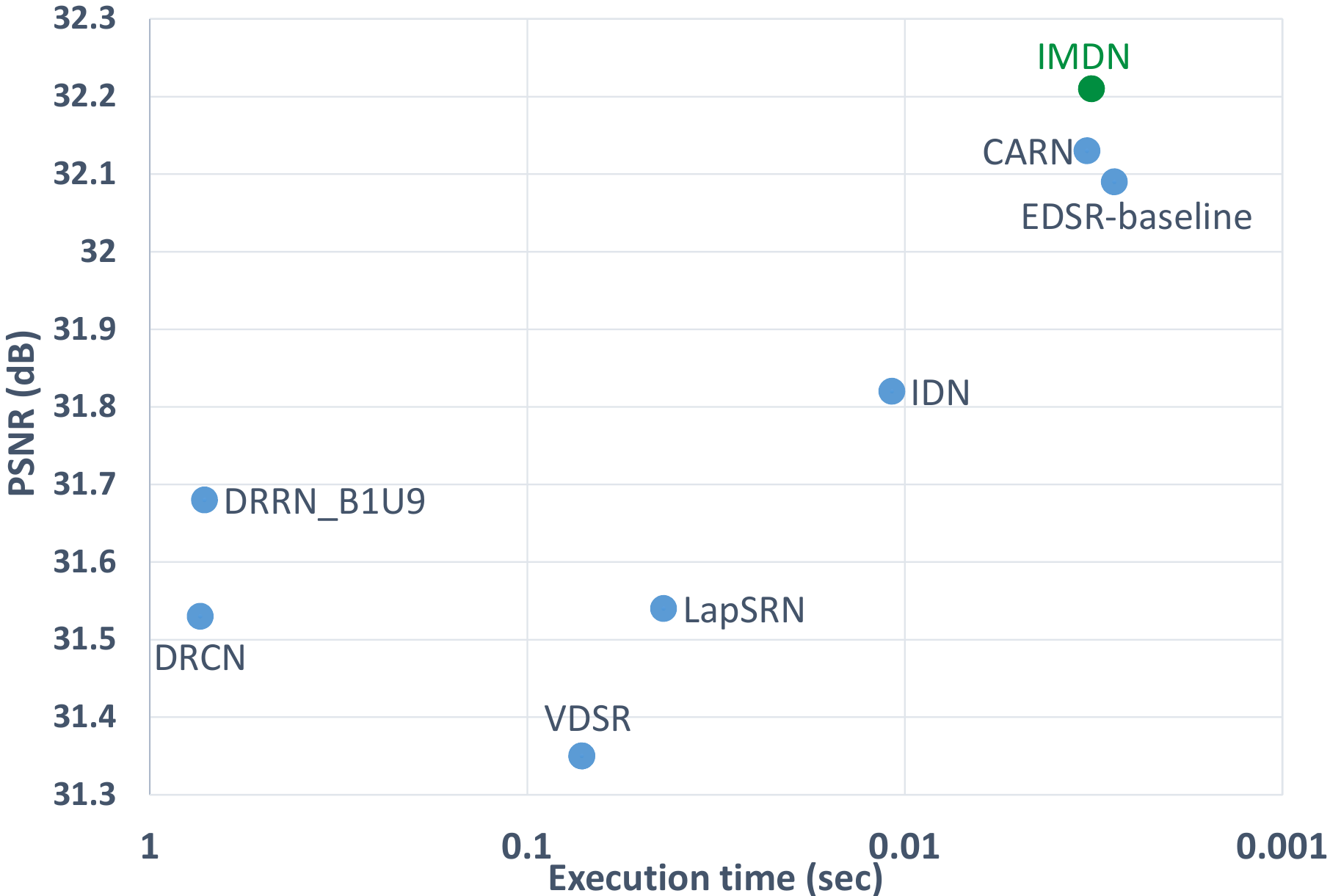}
	\vspace{-1mm}
	\caption{Trade-off between performance and running time on Set5 $\times 4$ dataset. VDSR, DRCN, and LapSRN were implemented by MatConvNet, while DRRN, and IDN employed Caffe package. The rest EDSR-baseline, CARN, and our IMDN utilized PyTorch.}
	\label{fig:time}
	\vspace{-1mm}
\end{figure}

We use official codes of the compared methods to test their running time in a feed-forward process. From Table~\ref{tab:memory-time}, we can be informed of actual execution time is related to the depth of networks. Although EDSR has a large number of parameters (43M), it runs very fast. The only drawback is that it takes up more graphics memory. The main reason should be the convolution computation for each layer are parallel. And RCAN has only 16M parameters, its depth is up to 415 and results in very slow inference speed. Compared with CARN~\cite{CARN} and EDSR-baseline~\cite{EDSR}, Our IMDN achieves dominant performance in term of memory usage and time consumption.

For more intuitive comparisons with other approaches, we provide the trade-off between the running time and performance on Set5 dataset for $\times 4$ SR in the Figure~\ref{fig:time}. It shows our IMDN gains comparable execution time and best PSNR value.

\section{Conclusion}\label{sec:conclusion}
In this paper, we propose an information multi-distillation network for lightweight and accurate single image super-resolution. We construct a progressive refinement module to extract hierarchical feature step-by-step. By cooperating with the proposed contrast-aware channel attention module, the SR performance is significantly and steadily improved. Additionally, we present the adaptive cropping strategy to solve the SR problem of an arbitrary scale factor, which is critical for the application of SR algorithms in the actual scenes. Numerous experiments have shown that the proposed method achieves a commendable balance between factors affecting practical use, including visual quality, execution speed, and memory consumption. In the future, this approach will be explored to facilitate other image restoration tasks such as image denoising and enhancement.

\begin{acks}
	This work was supported in part by the National Natural Science Foundation of China under Grant 61432014, 61772402, U1605252, 61671339 and 61871308, in part by the National Key Research and Development Program of China under Grant 2016QY01W0200, in part by National High-Level Talents Special Support Program of China under Grant CS31117200001.
\end{acks}

\clearpage
\bibliographystyle{ACM-Reference-Format}
\balance
\bibliography{sample-base}

\end{document}